\documentclass[usenatbib,usegraphicx]{mn2e}

\catcode`\@=11
\newcommand{\amst}{\ifmmode\,\mbox{\AA}\else$\,$\AA\fi}
\newcommand\persqcm{\ifmmode\,{\rm cm^{-2}}\else$\,\rm cm^{-2}$\fi }
\newcommand\percubcm{\ifmmode\,{\rm cm^{-3}}\else$\,\rm cm^{-3}$\fi }
\newcommand{\persecond}{\ifmmode\,{\rm s^{-1}}\else$\,\rm s^{-1}$\fi}
\newcommand{\cccs}{\ifmmode\,{\rm cm^3\,s^{-1}}\else%
$\,\rm cm^3\,s^{-1}$\fi}
\newcommand{\flu}{{\ifmmode\,{\rm erg\,cm^{-2}\,s^{-1}}\else%
$\rm\,erg\,cm^{-2}\,s^{-1}$\fi}}
\newcommand{\kms}{{\ifmmode\,{\rm km\,s^{-1}}\else%
$\m@th{\,{\rm km\,s^{-1}}}$\fi}}
\newcommand{\kK}{{\ifmmode\,{\rm kK}\else$\,$kK\fi}}
\newcommand{\K}{{\ifmmode\,{\rm K}\else$\,$K\fi}}
\newcommand{\yr}{{\ifmmode\,{\rm yr}\else$\,$yr\fi}}

\newcommand{\alfaef}{\ifmmode\alpha_{\rm eff}\else%
$\alpha_{\rm eff}$\fi}
\newcommand{\hb}{\ifmmode{\rm H\beta}\else$\rm H\beta$\fi}
\newcommand{\lam}{\ifmmode\lambda\,\else$\lambda\,$\fi}
\newcommand{\lala}{\ifmmode\lambda\lambda\,\else$\lambda\lambda\,$\fi}
\newcommand{\mult}[2]{$\rm #1$--$\rm #2$}
\newcommand{\teff}{\ifmmode T_{\rm eff}\else$T_{\rm eff}$\fi}
\newcommand{\term}[1]{\ifmmode{\rm #1}\else$\rm #1$\fi}
\newcommand{\oric}{\ifmmode\theta^1\,{\rm C~Ori}\else%
$\theta^1\,{\rm C~Ori}$\fi}

\newcommand{\hii}{\mbox{H\thinspace{\sc ii}}}
\newcommand{\hp}{\ifmmode{\rm H^+}\else H$^+$\fi}
\newcommand{\hei}{\mbox{He\thinspace{\sc i}}}
\newcommand{\nii}{\mbox{N\thinspace{\sc ii}}}
\newcommand{\np}{\ifmmode{\rm N^+}\else N$^+$\fi}
\newcommand{\npp}{\ifmmode{\rm N^{+2}}\else N$^{+2}$\fi}

\newcommand{\oiii}{\mbox{O\thinspace{\sc iii}}}

\newcommand{\beq}{\begin{equation}}
\newcommand{\eeq}{\end{equation}}
\newcommand{\ecua}[1]{equation~(\ref{#1})}
\newcommand{\Ecua}[1]{Equation~(\ref{#1})}
\newcommand{\clo}{{\sc cloudy}}
\newcommand{\nebu}{{\sc nebu}}
\newcommand{\wmb}{{\sc wmbasic}}

\defcitealias{baldwin00}{BVV}
\defcitealias{esteban04}{EPG}
\defcitealias{escvict}{EV}

\newcommand{\aap}{A\&A}
\newcommand{\aaps}{A\&A Suppl.}
\newcommand{\aj}{AJ}
\newcommand{\apj}{ApJ}
\newcommand{\apjl}{ApJ Letters}
\newcommand{\apjs}{ApJS}
\newcommand{\mnras}{MNRAS}
\newcommand{\pasp}{PASP}
\catcode`\@=12
\title[The NII spectrum of Orion]{The NII spectrum of 
the Orion Nebula}
\author[V. Escalante and C. Morisset]{Vladimir Escalante$^1$
and Christophe Morisset$^2$ 
\thanks{E--mail: v.escalante@astrosmo.unam.mx\hfill\break 
morisset@astroscu.unam.mx}\\ 
$^1$Centro de Radioastronom{\'\i}a y Astrof{\'\i}sica, Ap.~Postal 72--3, 
C.~P.~58091, Morelia, Michoac\'an, M\'exico\\ 
$^2$Instituto de Astronom{\'\i}a, Ap.~Postal 70--264, 
C.~P.~04510, M\'exico, DF, M\'exico}

\begin{document}

\pagerange{\pageref{firstpage}--\pageref{lastpage}}

\maketitle

\label{firstpage}

\begin{abstract}
The predicted emission spectrum of \nii\ is 
compared with observations of permitted lines in the 
Orion nebula. 
Conventional nebular models show that the 
intensities of the more intense lines 
can be explained by fluorescence of starlight 
absorption with a N abundance that is consistent 
with forbidden lines. 
Lines excited mostly by recombination on the other 
hand predict high N abundances.  
The effects of stellar and nebular parameters and of the 
atomic data on the predicted intensities are examined. 
\end{abstract}


\section{Introduction}

Recent deep spectroscopic surveys of the Orion nebula 
(Esteban et al.~1998; Baldwin et al.~2000, hereafter BVV; 
Esteban et al.~2004, hereafter EPG) show that \npp+e recombination 
cannot explain the intensities of the \nii\ 
permitted lines with a nitrogen abundance that is consistent with the 
\nii\ forbidden line intensities. 
\citet{seaton68} first suggested the possibility that permitted lines 
of C and O ions in nebulae may be excited by continuum fluorescence 
of starlight, 
and \citet{grandi76} noted that this also must be 
an important mechanism to excite the \nii\ lines in Orion. 
Grandi suggested that additional absorption of photons of the 
\hei\ \mult{1s^2\ ^1S_0}{1s8p\ ^1P_1^o} line at 
\lam508.643\amst\ from the diffuse field of the nebulae 
by the \nii\ \mult{2p^2\ ^3P_0}{2p4s\ ^3P_1^o} transition at 
\lam508.668\amst\ 
followed by decay to 3p terms would enhance the observed intensity 
of some lines. 
There is few direct evidence of the plausibility of this mechanism. 
\citetalias{baldwin00} and \citetalias{esteban04} observed 
lines at \lam3829.92, \lam3838.47 and \lam3856.27 that could be 
produced by the \mult{3p\ ^3P_1}{4s\ ^3P_2^o}, \mult{3p\ ^3P_2}{4s\ ^3P_2^o} 
and \mult{3p\ ^3P_2}{4s\ ^3P_1^o} transitions respectively. 
Unfortunately those lines are probably blended with lines 
from other elements, 
and their identification and intensity are uncertain. 
\citet{sharpee} have detected those and other lines 
of the \mult{3p\ ^3P}{4s\ ^3P^o} multiplet in a 
planetary nebula (IC~418), where the \nii\ spectrum is probably 
excited by fluorescence.  
The \term{4s\ ^3P_2^o} level requires pumping of the 
the \nii\ \mult{2p^2\ ^3P_2}{2p4s\ ^3P_2^o}\lam508.697\amst\ 
transition, 
which lies 32\kms\ from the \hei\ line. 
Other lines from the \term{4s\ ^3P^o} term like 
those of the \mult{3p\ ^3D}{4s\ ^3P^o} multiplet in the 
\lala3311.42--3331.31 interval, 
which should be as intense as the \mult{3p\ ^3P}{4s\ ^3P^o} multiplet, 
were not detected by EPG. 
The efficiency of this Bowen--type line fluorescence depends 
heavily on uncertain nebular parameters that are needed in the 
theory of line radiative transfer, 
and modeling can become quite arbitrary. 
Therefore we will not consider it in this 
work \citep[for further discussion see][]{escalante,liu01}. 

\begin{figure}
 \includegraphics[width=9.6cm]{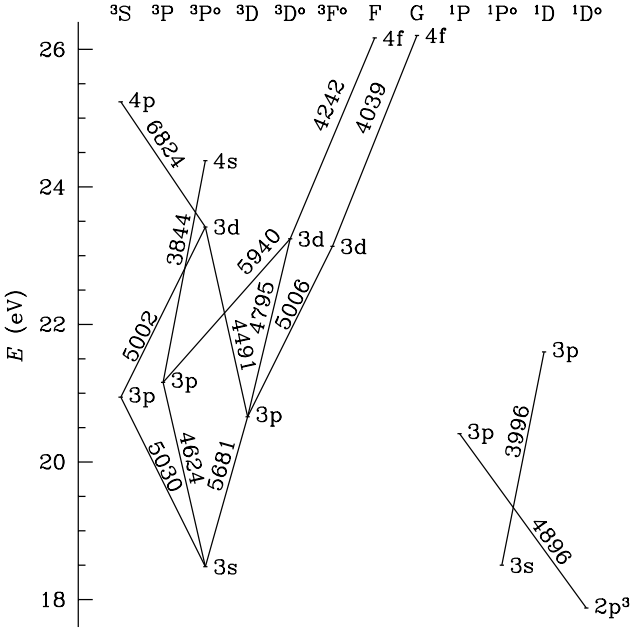}
 \caption{The singlet and triplet \nii\ systems. Terms 
  \term{3d\ ^3P^o}, \term{3d\ ^3D^o} and \term{4s\ ^3P^o} are 
  connected by resonant transitions to the \term{2p^2\ ^3P} ground 
  term at 0~eV. Wavelengths are term--averaged values in \amst. 
  }
  \label{grotn2}
\end{figure}

The critical parameters that determine the intensities of the lines 
and the relative importance of the fluorescence mechanism over 
the recombination process are the \np\ and \npp\ column densities in 
the gas and the stellar UV radiation field. 
Absorption of a UV photon by a resonant transition between a 
ground configuration state and an excited state has a higher 
probability of subsequent reemission in the same transition. 
Decay to an intermediate state will be favoured when the 
optical depth of the resonant transition is large, 
and the resonant photon is scattered a few times until it is 
converted into a lower energy photon producing a subordinate line. 
However the optical depth of the resonant transition 
must not become too large in order to allow 
enough resonant photons to penetrate into the \np\ zone. 
The efficiency of the continuum fluorescence excitation in 
\np\ depends on the transfer of resonant photons between 
the lowest resonant transition that can produce a subordinate 
line, \mult{2p^2\ ^3P}{3d\ ^3D^o}\lala533.51-533.88\amst, 
and the ionization limit at 419\amst. 
We have also included transitions from the 
\term{2p^2\ ^1D_2} and \term{^1S_0} metastable 
states--populated mostly by collisions--to 
other singlets in order to consider the observation of 
singlet lines in Orion. 
In the singlet system the lowest transition that produces a subordinate
line is \mult{2s^22p^2\ ^1S_0}{2s2p^3\ ^1P_1^o}\lam745.84\amst. 
Fig.~\ref{grotn2} shows the observed transitions in Orion of 
the singlet and triplet \nii\ systems. Possible observations 
of quintet lines and higher excitation states are discussed in 
section~\ref{fluorspec}. 

The observed intensities of permitted lines in planetary nebulae 
are often higher than their intensities predicted by 
recombination rates with CNO abundances measured from 
forbidden lines \citep[ and references therein]{liu95,liu01,luo}. 
Some of the \nii\ lines observed in PNe have 4f upper 
levels, which are excited mainly by recombination. 
In Orion there exists a similar situation and we will show 
that fluorescence cannot account for the excess intensity of 
lines from 4f levels. 
The accuracy of the recombination theory can be more easily 
tested by comparing line ratios from 4f decays because 
all the transitions involved are optically thin and 
the absolute line intensities have a similar dependence 
on temperature and density. 

The fluorescence theory is more difficult to test because 
it depends, often non--linearly, on several model parameters. 
We will use the ratio of predicted over observed intensities, 
averaged over all \nii\ lines with reasonably accurate identifications and 
measurements, hereafter referred to as 
$R=\langle I_{\rm cal}/I_{\rm obs}\rangle$, 
to test the model parameters. 
The scatter around $R$ will be used as a test of the atomic data 
and the details of the transferred stellar continuum. 

Realistic nebular models and hot star model atmospheres along 
with available atomic data bases 
can explain successfully the intensities of a majority of the 
forbidden lines in Orion as well as general observed 
trends of ion abundances in Galactic HII regions 
\citep[see for example][]{baldwin91,stasinska}. 
This paper demonstrates that the intensity of most of the 
\nii\ permitted lines in the Orion nebula can be predicted 
by fluorescence of the starlight continuum and some contribution of 
recombination by these models with currently accepted physical 
conditions and abundances of the nebula. 

%                                      generated with grotn2.sm

\section{Calculation of population densities}

\subsection{Atomic processes}
\label{atprocs}

The important processes that can produce excited states 
with low principal quantum numbers in \np\ at nebular 
temperatures are absorption of UV photons by 
transitions from ground and metastable states and recombinations 
of \npp. 
Subsequent decays of these states produce the optical \nii\ spectrum. 
UV radiation is provided mostly by the star \oric. 
The contribution of the diffuse continuum to the absorption is negligible. 

The number density of a \np\ excited state $j$ at 
a certain point in the nebula in steady state, $n_j$, is given by 
\beq
A_jn_j=\alpha_jn_en(\npp)+\sum_g\beta_{gj}n_g+\sum_{k>j}A_{kj}n_k\ ,
\label{popul}
\eeq
where $A_j=\sum_iA_{ji}$ is the spontaneous decay rate, 
$\alpha_j$ is the recombination (radiative plus 
dielectronic) coefficient to state $j$, $n_g$ is the population 
density of one of the five \np\ ground and metastable states, 
\term{2p^2\ ^3P_{0,1,2}, ^1D_2, ^1S_0}, and $\beta_{gj}$ is the 
radiation absorption rate. 
If $\bar J_\nu$ is the local starlight mean intensity 
averaged over the absorption profile, 
\beq
\beta_{gj}=\sigma_{gj}\left(4\pi \bar J_\nu\over h\nu\right)\gamma_{gj}\ ,
\label{beta}
\eeq
where $\sigma_{gj}=0.02654f_{gj} \,{\rm cm^2\,Hz}$ is the absorption
cross section and $f_{gj}$ is the f--value of transition $g\to j$. 
$\bar J_\nu$ is attenuated by continuum opacity and geometry. 
The ``pumping probability'' to account for the attenuation 
due to the transition $g\to j$ as defined by \citet{ferland} is 
\beq
\gamma_{gj}=\int e^{-\tau_0\phi(x)}\phi(x)\,dx\ ,
\label{pumprob}
\eeq
where $\phi(x)$ is the normalized Voigt profile for a displacement 
$x=(\nu-\nu_0)/\Delta \nu$, 
and $\tau_0=\sigma_{gj}N_g/\Delta\nu$ 
is the mean optical depth for a column density 
$N_g=\int n_g\,dr$ integrated along a ray from the star. 

An actual stellar spectrum shows absorption lines 
superposed on the continuum and the P--Cygni profiles of 
the wind. 
If the structure of the stellar absorption spectrum varies 
in scales comparable to the width of $\Delta\lam\sim0.01\amst$ 
of the UV resonant lines in the nebular gas, 
\ecua{pumprob} should be changed to 
\beq
\gamma_{gj}=\int \psi(x) e^{-\tau_0\phi(x)}\phi(x)\,dx\ ,
\label{pumprob2}
\eeq
where $\psi(x)=J_\nu/\bar J_\nu$ is the continuum stellar profile 
around the resonant line. 
Grids with that resolution are available in the 
optical \citep{murphy}. 
In the UV \citet{smith} and \citet{amiel} have published 
models with a lower resolution. 

The populations of the metastable states at nebular temperatures 
are controlled by electron collisions, 
and are nearly independent of other excited states. 
The system of equations~(\ref{popul}) is triangular, 
and can be solved in terms of the cascade matrix, 
$C_{kj}$ \citep{seaton59}. 
$C_{kj}$ is the probability that a state $j$ is produced 
by the excitation of state $k$ followed by 
radiative decays by all possible routes ending in $j$. 
\Ecua{popul} thus becomes 
\beq
A_jn_j=\alpha^{\rm eff}_jn_en(\npp)+\sum_g\beta^{\rm eff}_{gj}n_g\ ,
\label{popj}
\eeq
where
\begin{eqnarray}
\alpha^{\rm eff}_j=\sum_{k\ge j}\alpha_kC_{kj}\ ,
\label{alfaef} \\
\beta^{\rm eff}_{gj}=\sum_{k\ge j}\beta_{gk}C_{kj}\ ,
\label{betaef}
\end{eqnarray}
are the effective recombination coefficient and the 
effective fluorescence rate respectively. 
In practice the sums in equations~(\ref{alfaef}) 
and~(\ref{betaef}) must be truncated and a 
correction must be added to \ecua{alfaef} by extrapolation 
or by using hydrogenic populations (Escalante \& Victor 1990, 
hereafter EV) 
because recombination coefficients of individual levels 
decrease slowly with 
$n$ and the contribution of states with high angular momentum 
and high $n$ must be included. 
Care must also be taken to avoid roundoff errors in the summations. 
The effective fluorescence rate in~\ecua{betaef} is 
less sensitive to that type of correction because the 
absorption rate $\beta_{gk}$ is non--zero only for states 
connected to the ground and metastable states by dipole--allowed transitions,  
and decreases more rapidly with the 
principal quantum number $n$ of the upper state. 
The contribution of levels with $4\le n\le 12$ and orbital 
angular momentum $0\le l\le 2$ in \ecua{betaef} 
is less than 5\% for the \nii\ lines observed in Orion. 
The contribution of f states to the fluorescence rate is even less important 
because they are not connected by resonant transitions to the 
ground term, 
and the transitions $n{\rm d}\to m{\rm f}$ have a low relative probability. 
By eliminating states with $n>9$ and $l>2$ 
in \ecua{betaef}, the CPU time decreases by a factor of 10. 
However the f states (and higher $l$ states) must be included in 
the cascade due to recombinations, which tend to favor 
high--angular momentum states.  

Most of the observed \nii\ transitions in Orion come from decays of the 
3p and 3d triplet terms. Practically all the excitations of 
the 3d terms are produced by direct absorptions in the multiplets 
\mult{2s^22p^2\ ^3P}{2p3d\ ^3P^o}\lala529.36--529.87~\AA\ and 
\mult{2s^22p^2\ ^3P}{2p3d\ ^3D^o}\lala533.51--533.88~\AA. 
Cascades contribute negligibly to the populations of the 3d terms. 
The 3p terms, 
not being connected by direct transitions to the ground term, 
have more varied excitation channels. 
Between half and 80\% of excitations of 
3p terms come from decays of 3d terms. 
The rest comes mostly from absorptions in the multiplet 
\mult{2s^22p^2\ ^3P}{2p4s\ ^3P^o}\lam\lam508.48--509.01~\AA\  
and higher s and d states. 
Table~\ref{tabfvals} shows a comparison of $f$--values for these multiplets, 
which are critical in our calculations. 

%         comparison of f values
\begin{table}
  \caption{$f$ absorption values for some \nii\ resonant multiplets.}
  \label{tabfvals}
  \begin{tabular}{@{}crllll}
    \hline
    Multiplet&$\lam$(\AA)&\multicolumn{4}{c}{$f$}\\
    &\multicolumn{1}{c}{(1)}&\multicolumn{1}{c}{(2)}
    &\multicolumn{1}{c}{(3)}&\multicolumn{1}{c}{(4)}
    &\multicolumn{1}{c}{(5)}\\
    \hline
    \mult{2p^2\ ^3P}{4s\ ^3P^o}&508.74&0.00987&0.0175&0.0105&0.0124\\
    \mult{2p^2\ ^3P}{3d\ ^3P^o}&529.68&0.102&0.175&0.103&0.138\\
    \mult{2p^2\ ^3P}{3d\ ^3D^o}&533.67&0.294&0.575&0.300&0.378\\
    \hline
  \end{tabular}

  \medskip
  (1) Multiplet--averaged wavelength.\\
  (2) \citet{wiese}.\\
  (3) \citet{victor}.\\
  (4) \citet{opacity} (Opacity Project).\\
  (5) \citet{nussb}.\\
\end{table}

The advantage of using the cascade matrix is that it 
needs to be computed once in the model in either 
the optically thin case (case A) or the extreme thick case (case B) 
because it depends solely on the Einstein coefficients, 
or more precisely, on the branching ratios $P_{kj}=A_{kj}/\sum_iA_{ki}$. 
In the escape probability formalism the equations must 
be modified by substituting the Einstein coefficients of 
transitions connected to the states of the ground 
configuration \term{2p^2\ ^3P,\ ^1D,\ ^1S} by 
$P_eA_{kg}$, where $P_e$ is the escape probability to be discussed 
below. 
$P_e$ is a local quantity that depends on the 
optical depth of the line as a function of position and thus the 
cascade matrix must be recomputed at each point in the nebula. 
However only a small fraction of the matrix elements depend on $P_e$. 
We found that recomputing only this fraction of the 
cascade matrix reduces the CPU time by a factor of 30. 

\subsection{Escape probabilities}

We have used the escape probability theory to handle
nearly 200 resonant transitions that produce 2630
transitions by fluorescence. 
A review of the limitations of
this theory has been given by \citet{dumont}, 
but a full line transfer calculation is beyond present 
computation capabilities.
There are many approximations for the escape 
probability function, 
which differ by several factors at high optical depths. 
In an ionized nebula absorption of resonant photons by an overlapping 
continuum is important \citep{hummer}. 
In a uniform slab of total optical thickness $T$ 
the escape probability at depth $\tau$ is 
$P_e=bF(b)+(1/2)[K_2(\tau,b)+ K_2(T-\tau,b)]$, 
where the functions $F$ and $K_2$ are 
defined in \citet{humstorey}, 
and $b=k_c/k_l$ is the continuum--to--mean line opacity ratio. 
The term
\beq
bF(b)=\int_\infty^\infty {k_c\over k_l+k_c}\phi(x)\,dx\ ,
\label{bfhumm}
\eeq
is the probability that the resonant photon will be lost by 
continuum absorption. 
The effect of $P_e$ is to increase considerably the 
probability that a resonant absorption decays into the subordinate 
line rather than reemitting the resonant photon. 
The average ratio, 
$R=\langle I_{\rm cal}/I_{\rm obs}\rangle$, 
increases by a factor of 20 between the two limits 
$P_e\equiv 1$ (case~A) and $P_e\equiv 0$ (case~B) respectively. 
A spectrum dominated by fluorescence, 
however, must be in an intermediate regime in order 
to allow the penetration of UV stellar photons through a large column density 
of absorbers at the same time that the resonant photons are 
scattered repeatedly. 

All the resonant photons that produce the \nii\ optical spectrum 
have a high probability of being lost by conversion into other 
lines besides being absorbed by the continuum, and the probability 
of a large number of scatterings is small. Consequently we used a Doppler 
core in the calculation of $K_2(\tau,b)$ and $F(b)$. 

\subsection{Atomic Data}

%         comparison of fluorescence efficiencies
\begin{table}
  \caption{Effective fluorescence rates for some \nii\ lines.}
  \label{tabbetas}
  \begin{tabular}{@{}crllll}
    \hline
    Line&$\lam$(\AA)&\multicolumn{4}{c}{$P_{ji}\sum_g\beta^{\rm eff}_{gj}
      \ (10^{-8}\persecond)$}\\\\
    \mult{i\ \ \ }{\ \ \ j}&&\multicolumn{1}{c}{(1)}&\multicolumn{1}{c}{(2)}
    &\multicolumn{1}{c}{(3)}&\multicolumn{1}{c}{(4)}\\
    \hline
    \mult{3s\ ^3P^o_2}{3p\ ^3P_2}&4630.5&10.8&25.7&16.1&13.7\\
    \mult{3p\ ^3D_3}{3d\ ^3D^o_3}&4803.3&4.6&4.5&4.5&4.6\\
    \mult{3s\ ^3P^o_2}{3p\ ^3D_3}&5679.6&7.1&16.3&10.3&8.9\\
    \mult{3p\ ^3P_2}{3d\ ^3D^o_3}&5941.7&8.0&8.6&8.6&8.0\\
    \hline
  \end{tabular}

  \medskip
  (1) Only \citet{wiese}.\\
  (2) Only \citet{victor}.\\
  (3) Same as (2) plus doubly--excited configurations\\
  and $\Delta S\ne0$ transitions.\\ 
  (4) All transitions included.\\
\end{table}

The main source of $A$-- and $f$--values for this work is the compilation 
of \citet{wiese}, which is based primarily on the Opacity 
Project data \citep{opacity} and configuration--interaction calculations, 
but also includes intermediate coupling calculations and 
experimental measurements. 
That compilation has data for most lines of the series 
${\rm 2s^22p(^2P)}nl$ with 
$n\leq 5$ and $l\leq 2$, and ${\rm 2s2p^2(^4P)}nl$ with 
$n\leq 2$ and $l\leq 1$, including spin--forbidden transitions 
between quintet and triplet terms. 
The model potential calculations from \citet{victor} 
were used for the rest of the transitions needed in the cascade 
matrix of the effective fluorescence rate in \ecua{betaef}. 
Transition probabilities between fine structure levels were 
obtained by applying LS fractions to the multiplet data \citep{allen}. 

Table~\ref{tabbetas} shows the effect of different atomic databases on 
the effective fluorescence rate~(\ref{betaef}) summed over the 
ground and metastable states times the 
branching ratio of the subordinate line. 
The continuum is a blackbody spectrum at $T=37\kK$ with a photon flux of 
$\rm 9.24\times10^{-4}\persqcm\persecond\,Hz^{-1}$ at \lam533.7\amst\ 
and $\tau_0=0$ for all lines, i.e., $\gamma_{gj}=1$, $P_e=1$. 
The compilation of \citet{wiese} does not include transitions to 
levels with principal quantum number greater than $5$ and produce 
lower fluorescence rates of lines with upper p levels because 
those levels have cascade contributions from higher levels. 
The model potential data of \citet{victor} does not have 
transitions to doubly--excited configurations and consequently give 
higher rates as shown in the fourth column of table~\ref{tabbetas}. 
Transitions between states with the \term{2s2p^3} configuration 
and the singly excited configurations \term{2s^22p3p} can change 
the branching ratios significantly. 
The entries in the fifth column 
have been complemented with transitions to states with 
the \term{2s2p^2} core configuration and spin-forbidden transitions 
taken from \citet{wiese}. 
The last column combines both data sets. 
The cascade matrix elements tend to be similar for different data sets 
because systematic differences in the atomic parameters between 
different data sets cancel out in the branching ratios 
$P_{ji}=A_{ji}/A_j$. 

A recent calculation of effective recombination coefficients 
by \citet{kisi} shows a general agreement with 
the model potential calculations of \citet{escvict} (hereafter EV) 
and the calculations of \citet{pequignot} for the 
\term{2s^22p(^2P^o)3d} and 4f configurations. 
The most important differences between the three calculations 
are in the branching ratios of multiplets from the 
\term{2s^22p(^2P^o)3p\ ^3D} term. 
The accuracy of the model potential in this case was limited by the lack 
of observed energies in the \term{2s^22p(^2P^o)np} series. 
The best agreement between the three data sets is in the 4f terms, 
where most of the contribution to the 
recombination comes from levels with small non-hydrogenic effects. 
The main uncertainty is in the line fractions involving 
4f terms, where LS--coupling is not a good approximation. 
\citetalias{escvict} 
used LK coupling for these terms, and line fractions for other 
couplings are available \citep{escgong}, but general 
intermediate--coupling calculations for those states are clearly needed. 

This work uses the recombination coefficients of \citetalias{escvict} 
with branching ratios given by the A--values of \citet{victor} 
and \citet{wiese}. 
Effective recombination coefficients for the levels of a 
term were obtained by assuming that the coefficients are proportional to 
the statistical weights of the levels. 

\section{Model calculations}

\subsection{Nebular models}

In order to determine the electron, \npp, and \np\ densities 
in \ecua{popj}, as well as the temperature and opacity at each point in 
the nebula, 
we used the codes \clo\ \citep[ version 90.05]{cloudy} and 
\nebu\ \citep{pequignot01,morisset}. 
Models of the Orion nebula support the existence of a main emitting 
layer at the back of a cavity in the OMC--1 molecular cloud.  
The thickness of the layer is highly variable across the nebula \citep{wen}. 
We approximated the layer by a plane--parallel model at constant pressure 
with \clo\ and constant density with \nebu. 
\clo\ allowed us to use the grain composition used by \citet{baldwin91} 
in their Orion model. 
Predicted line intensities by \clo\ show some sensitivity when 
the radiation is included in the pressure law \citep{baldwin96}. 
Models with a constant gas pressure produce larger \np\ column 
densities than models with a constant gas plus radiation pressure. 
We have not tried to find a single model fit to the observed 
forbidden line intensities in Orion. 
Instead we have run a series of models to find the most important 
dependencies of the \nii\ lines excited by fluorescence on model 
parameters. 

\subsection{Model atmospheres}

The Orion nebula is mostly excited by \oric. 
The other Trapezium stars increase the fluorescence of 
the \nii\ lines by less than 2\% and will not be considered 
in this calculation. 
\oric\ is a class V star with variable wind features 
that produce uncertainties in the determination of 
its spectral classification and effective temperature. 
Different authors give values close to $\teff=39\kK$ and 
$\log(g)=4$ for this star \citep[ and references therein]{howarth,hillen}. 
However comparisons of the intensities of forbidden lines 
with nebular model predictions suggest temperatures as low as 
36\kK\ \citepalias{baldwin00}. 
Metalicity measurements by \citet{cunha} show that the 
Trapezium stars are slightly underabundant with respect to the Sun. 

The calculation of the fluorescence of the gas needs a high resolution 
stellar spectrum. 
This is important with hot massive stars, 
which have expanding atmospheres with a dense forest of absorption lines 
and broad overlapping P Cygni profiles. 
The emission peaks and absorption troughs of the overlapping 
P Cygni profiles can 
increase or decrease the absorption rate in \ecua{pumprob} 
by large factors and change the fluorescence excitation rate 
significantly. 
On the other hand nebular models usually smooth 
the spectrum of the exciting star to calculate the ionization structure. 
In order to take into account the detailed spectral structure of the 
model atmosphere, 
we input the unattenuated stellar spectrum into the nebular 
model and extracted the predicted attenuated local continuum 
at each point in the nebula. 
To calculate the pumping rate in \ecua{pumprob}, 
the full resolution spectrum was read and interpolated at the 
absorption frequency and was scaled by the ratio of the 
attenuated to the unattenuated stellar continuum predicted by the 
nebular model. 

The motion of the star with respect to the gas introduces a 
Doppler shift that can change the intensity of the 
continuum at the absorbing frequencies. 
The proper motion velocity of \oric\ is uncertain \citep{wen}. 
Doppler shifts of up to $\pm30\kms$ did not change predicted line 
intensities by more than a few percent with the resolution 
of about $\Delta\lam\sim1\amst$ in the far UV of the model atmospheres 
that we used. 
Therefore we assumed a static star with respect to the gas. 

Recent model atmospheres of O stars include the effects of line 
blanketing and line blocking of the stellar wind. 
We used model atmospheres calculated with the \wmb\ code
\citep{pauldrach,amiel} to account for these effects. 
We also used the LTE, line--blanketed atmospheres 
of \citet{kuru} for comparison purposes. 

\section{Basic model}

All transitions of the observed permitted lines in Orion 
end in excited states and are optically thin. 
Their intensities can thus be obtained by integration 
of the emissivity along the line of sight: 
\beq
I={h\nu\over4\pi}\int A_{ji}n_j\,dr
\label{intens}
\eeq
We now examine the dependence of the fluorescence
excitation on the stellar spectrum and the density and compare 
them with the observations of 
\citetalias{baldwin00} and \citetalias{esteban04}.  

We have adopted a set of central values for the parameters of 
the models that are given in table~\ref{tabmodpar}. 
These values are close to the ones recommended by \citet{baldwin91}, 
and we will refer to them as the basic model (BM). 
We discuss below a few features of this model. 

%         Table with model parameters 
\begin{table}
 \centering
  \caption{Model parameters for the basic model (BM)}
  \label{tabmodpar}
  \begin{tabular}{@{}ll@{\hsize=180pt}ll}
  \hline
    \teff*&37\kK                          &$n$ &$10^4\percubcm$\\
    $\log g$       &4.0                   &He  &0.095\\
    $Z$            &$1Z_\odot$            &C   &$3\times10^{-4}$\\
    $\phi_0$*      &$10^{12.9}\persecond$ &N*  &$6\times10^{-5}$\\
    $\theta$       &$0^{\rm o}$           &O   &$4\times10^{-4}$\\
    $V_{\rm turbulent}$&$8\kms$           &Ne  &$6\times10^{-5}$\\
  \hline
\end{tabular}

\medskip
* Variable parameters in this work. 
\end{table}

\subsection{Gas abundances and density.}
\label{gasabun}

%     From   /home/vladimir/atom/emis/n2/recomfrac.sxc
%         lines from 4f levels
\begin{table*}
  \caption{Intensities of lines from 4f levels in Orion and PNe 
     [$I(\nii\lam5679.56)=1$].}
  \label{tabflevs}
  \begin{tabular}{@{}cclrcccccc}
    \hline
    LK Multiplet&Line&\multicolumn{1}{c}{\lam(\AA)}
    &\multicolumn{1}{c}{\%}
    &\multicolumn{1}{c}{Theory}
    &\multicolumn{2}{c}{Orion}
    &\multicolumn{3}{c}{PNe}\\
    \mult{3d\ ^3L^o_J}{4f\ L'[K]_{J'}}&\mult{J}{[K]J'}
    &\mult{J}{[K]J'}&\multicolumn{1}{c}{(1)}&\multicolumn{1}{c}{(2)}
    &\multicolumn{1}{c}{(3)}&\multicolumn{1}{c}{(4)}
    &\multicolumn{1}{c}{(5)}&\multicolumn{1}{c}{(6)}
    &\multicolumn{1}{c}{(7)}\\
    \hline
    \mult{3d\ ^3F^o_J}{4f\ G[K]_{J'}}
    &\mult{3}{[{9/2}]4}
    &4026.08&13.9&0.24&*&*&*&*&*\\
    &\mult{2}{[{7/2}]3}
    &4035.08&23.8&0.41&--&--&0.24&0.23&0.32\\
    &\mult{4}{[{9/2}]5}
    &4041.31&40.7&0.70&--&0.30&0.58&0.53&0.58\\
    &\mult{3}{[{7/2}]4}
    &4043.53&17.4&0.30&--&--&0.35&0.21&0.23\\
    \mult{3d\ ^3D^o_J}{4f\ G[K]_{J'}}
    &\mult{3}{[{7/2}]3}
    &4201.35&--&--&--&0.14&--&--&--\\
    $\mbox{\mult{3d\ ^3D^o_J}{4f\ F[K]_{J'}}}\atop
    \mbox{}$
    &$\mbox{\mult{1}{[{5/2}]2}}\atop
    \mbox{\mult{2}{[{7/2}]3}}$
    &$\mbox{4236.93}\atop\mbox{4237.05}$
    &$\mbox{20.0}\atop\mbox{12.7}$
    &$\left.\mbox{0.20}\atop\mbox{0.13}\right\}$
    &0.23&0.16
    &$\left\{\mbox{0.17}\atop\mbox{0.26}\right.$
    &$\mbox{0.13}\atop\mbox{0.20}$ 
    &$\mbox{0.16}\atop\mbox{0.23}$\\
    &\mult{2}{[{5/2}]2}
    &4241.24&3.7&0.04&--&--&0.04&0.04&0.04\\
    &$\mbox{\mult{2}{[{5/2}]3}}\atop
    \mbox{\mult{3}{[{7/2}]4}}$
    &$\mbox{4241.76}\atop\mbox{4241.79}$
    &$\mbox{16.9}\atop\mbox{42.9}$
    &$\left.\mbox{0.17}\atop\mbox{0.42}\right\}$
    &0.20&0.28&0.49&0.54&0.51\\
    &\mult{3}{[{7/2}]3}
    &4242.49&1.6&0.02&--&0.28&--&--&--\\
    \hline
  \end{tabular}

\medskip
* Blended with \hei.\\
(1) Line fraction in LK coupling.\\
(2) Calculated from recombination rates at $T=10^4\K$ 
   in case A, \citetalias{escvict}.\\
(3) \citetalias{baldwin00}.\\
(4) \citetalias{esteban04}.\\
(5) NGC 6153, \citet{liu00}, intensities from scanned spectrum.\\
(6) M 1--42, \citet{liu01}.\\
(7) M 2--36, \citet{liu01}.\\
\end{table*}

The \npp\ abundance in the nebula can be estimated 
from the measurements of lines with upper 4f levels, 
which are populated mostly by recombination.
Table~\ref{tabflevs} lists the observed intensities of
these lines in Orion and their predicted recombination emission rates 
normalized to the observed intensities and recombination rate of the
\mult{3s\ ^3P^o_2}{3p\ ^3D_3}\lam5679.56\amst\ line.
\citetalias{baldwin00} and \citetalias{esteban04} measured the 
lines at \lala4236.91, 4237.05 and 4241.784, 
which account for  
92\% of the total strength of the \mult{3d\ ^3D^o}{4f\ F} multiplet  
if the 4f levels are 
described by an LK coupling scheme as shown in the table. 
We do not include in the abundance estimation 
the \lam4242.49 line because its measured 
intensity is much higher than 
the one predicted from the recombination theory.  
At $T_e=10^4\K$ the effective recombination coefficient of multiplet 
\mult{3d\ ^3D^o}{4f\ F} is $5.6\times10^{-14}\cccs$ \citepalias{escvict},  
which implies $\npp/\hp=9.5\times10^{-5}$ if the lines were 
produced solely by recombination. 
This abundance is higher than the one implied by the forbidden 
lines $\approx6\times10^{-5}$ in Orion and many galactic \hii\ regions 
\citep[e.g.,][]{shaver,affler}. 

We notice that \citetalias{esteban04} detected one 
line from the \mult{3d\ ^3F^o}{4f\ G} multiplet (\lam4041.31) 
although other lines of this multiplet theoretically should be 
more intense than the observed lines of the 
\mult{3d\ ^3D^o}{4f\ F} multiplet. 
The corresponding $\npp/\hp$ abundance from this line is 
$8.5\times10^{-5}$. 
\citet{liu01} has reported a \npp\ abundance of $4.47\times10^{-5}$ 
in Orion from the \lala4041.31 and 4043.53 lines, which is more 
consistent with the abundance from the forbidden lines. 

\begin{figure}
 \includegraphics[width=8.8cm]{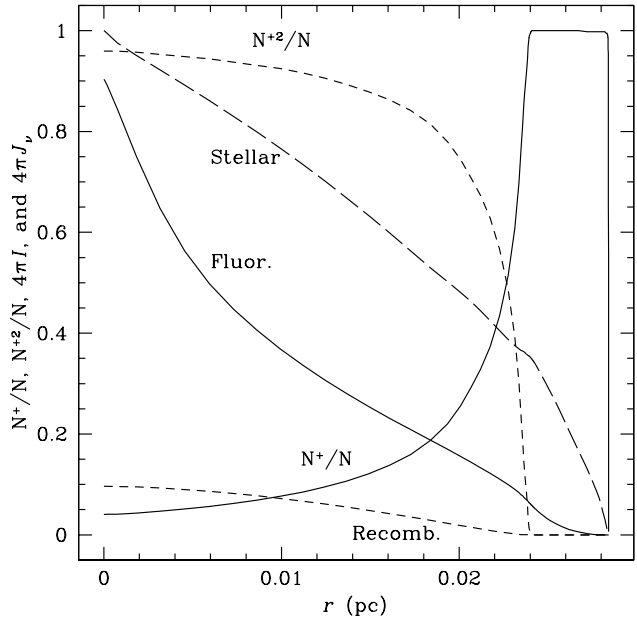}
 \caption{Ion fractions of \np, \npp, fluorescence  
  and recombination contribution to the intensity of the 
  5941.65\amst\ line and stellar continuum (long dash) at 533.73\amst\  
  as functions of geometrical depth in parsecs. 
  The intensities are normalized to the total intensity of the 
  line, $4\pi I=5.14\times10^{-4}\flu$ and to the stellar 
  flux on the illuminated side of the nebula, 
  $4\pi\bar J_\nu=1.402\times10^{-3}\flu\,{\rm Hz^{-1}}$.
  Earth and \oric\ are to the left. 
  }
  \label{emisvsr}
\end{figure}

Several authors \citep{liu00,liu01,tsamis,peimbert} have detected 
\nii\ lines in planetary nebula with relative rates that are 
more compatible 
with the recombination theory and will be the subject of future work. 
In a typical \hii\ region there is not enough \npp\ column density 
to produce a \np\ recombination spectrum. 
Fig.~\ref{emisvsr} shows separately the intensities 
given by \ecua{intens} (integrated from right to left) 
due to recombination and 
fluorescence as a function of depth for a line. 
While the recombination intensity grows linearly with distance 
the fluorescence intensity grows more rapidly. 
The fluorescence excitation is favoured by the more intense starlight 
continuum and low opacity in the near side of the nebula, 
while the recombination emissivity is more uniformly distributed. 
In the far side of the nebula the high optical depth in a 
resonant line scatters more photons and increases the probability 
of reabsorption in the line, 
thus increasing the pumping of the fluorescence emission. 
The measurements by \citetalias{baldwin00} show that most \nii\ permitted 
lines are blueshifted by 2 or 3~\kms\ with respect to the 
\nii\ forbidden lines, 
and support the idea that they are formed in different 
layers of the nebula. 

%                                      generated with emis.sm

We have chosen as a starting point the set of gas abundances of 
\citet{baldwin96}, 
which were derived to fit forbidden line intensities with \clo. 
The only exception was nitrogen for which we took a 
value of $6\times10^{-5}$ as an average from the determinations 
of \citet{baldwin96} and \citetalias{esteban04}. 
Differences in the abundances of the other elements between the
two sets are not important in \nii\ permitted line intensities.

The \np\ column density remains remarkably constant with varying 
gas density for a given effective temperature of the star, \teff. 
For the $\teff=37\kK$, $\log g=4$ atmosphere, 
$N(\np)=2.5\times10^{16}\persqcm$ for gas densities 
between 4,000 and 25,000\percubcm. 
Therefore we will adopt a fixed density of $n=10^4\percubcm$. 

\subsection{Stellar continuum and geometry}

The blister model, 
in which \oric\ is near the wall of the OMC--1 molecular
cloud \citep{zuckerman,balick}, 
gives the most likely geometry for the nebula. 
The Lyman photon flux of the star at a distance $r_0$ from the 
illuminated face of the nebula is $\phi_0=Q_0/4\pi r_0^2$, 
and can be constrained 
by the \hb\ observed brightness up to a geometrical factor that 
depends on the angle of illumination $\theta$ of the slab of gas 
as described by \citet{wen}. 
The value $\phi=10^{12.9}\persqcm\persecond$ and $\theta=0$, 
fixes the \hb\ intensity around 
$4\pi I=2.6\flu$ observed by \citet{baldwin91} 
if corrections are considered to account for reflected optical and 
absorbed UV radiations \citep{ferland01}. 
The \teff=37kK\ is too low for the probable spectral type of \oric, 
but was chosen because it reproduces the absolute intensities 
of the fluorescence lines with respect to the \hb\ flux and the 
N abundance measured from forbidden lines as discussed in 
section~\ref{stellartemp}.  

\subsection{The recombination spectrum.}

The 4f levels in \np\ are in an intermediate coupling between 
LK and jK couplings \citep{cowan}. 
In both couplings the total angular momentum is 
$J=K\pm1/2$. 
The line fractions for either coupling can 
be obtained from \citet{escgong}. 
We employed an LK classification to use the 
recombination rates of \citetalias{escvict}.  

Table~\ref{tabflevs} shows that all but two of the 
observed ratios of intensities of lines from 4f levels with 
respect to lines from 3d levels tend to be 
twice as strong and are closer to their 
predicted values by the recombination theory in PNe than in Orion.  
The two exceptions are the \lam4242.49 line, which seems 
much stronger than its predicted recombination rate, and 
the line of the \mult{3d\ ^3D^o}{4f\ G[7/2]} multiplet, which  
violates the $\Delta L=0,\pm1$ selection rule in LK 
coupling due to strong mixing with the \term{4f\ F[7/2]} term 
\citep{cowan}, and has no predicted recombination rate by 
\citetalias{escvict}. 
\citet{liu00,liu01} observed lines from other 
4f terms in PNe, which have not been detected in Orion and are 
not listed in table~\ref{tabflevs}. 
The higher intensity of the lines from 4f levels with 
respect to the lines from 3d levels in PNe indicates that 
fluorescence is less important relative to recombination 
in these objects because recombination can efficiently 
populate levels with high angular momentum 
while fluorescence is limited to resonant levels. 
One important exception to this trend is PN IC~418, 
which shows a strong enhancement of lines from 3p and 3d levels 
with respect to lines from the 4f levels \citep{sharpee}. 
This nebula has a lower ionization level than most PNe that 
favors the excitation of \nii\ lines by fluorescence over 
recombination as in Orion, and will be the subject of a 
forthcoming paper. 

Fluorescence can only contribute to population of the 4f levels 
through transitions from higher d levels, 
but this process is inefficient because 
the transition probabilities are much smaller 
than transitions to p levels. 
For example, 
fluorescence can populate the \term{4f\ F} levels through absorptions 
in the multiplet \mult{2p^2\ ^3P}{5d\ ^3D^o}\lam453,  
but in a pure case B 
the \term{5d\ ^3D^o} term decays to the terms \term{3p\ ^3P},  
\term{3p\ ^3D}, \term{2s2p^2(^4P)3s\ ^3P}, 
\term{4p\ ^3P} and \term{4f\ F} among others  
with branching ratios 
$P({\rm 5d,3p\ ^3P})=0.70$, 
$P({\rm 5d,3p\ ^3D})=0.14$, 
$P({\rm 5d,3s})=0.09$, 
and $P({\rm 5d,4f})=0.04$. 
As shown in section~\ref{gasabun}, 
\npp\ abundances obtained from 
recombination rates of the 4f lines in Orion are 
1.6 times the N abundance derived from collisionally excited lines. 
Suppose that the intensity of a 3d--4f line were enhanced 
with respect to \hb\ by fluorescence populating the 5d term 
with rate $B({\rm 5d})$: 
\beq
{I_\lam\over I_\hb}={4861\over\lam}\,P(\lam){E+P(5d,4f)B({\rm 5d})\over
E(\hb)}
\eeq 
where $E(\hb)=\int\alpha^{\rm eff}_\hb n(\hp)n_e\,d\ell$,  
$E=\int\alpha^{\rm eff}_{\rm 4f\,F}n(\npp)n_e\,d\ell$ and 
$P(\lam)$ is the branching ratio of the 3d--4f line with 
\lam\ given in \amst. 
To account for the excess abundance of 0.6 
obtained from the effective recombination rate of 
multiplet \mult{3d\ ^3D^o}{4f\ F}\lala4236.91--4247.20, 
we need $B({\rm 5d})=0.6E/P({\rm 5d,4f})$,  
which means a rate of population of the \term{3p\ ^3P} term of 
$B({\rm 5d})P({\rm 5d,3p\ ^3P})=10.5E$. 
The contribution to the intensity of a multiplet like 
\mult{3s\ ^3P}{3p\ ^3P}\lala4601.48--4643.08 
due to this additional excitation with 
$\npp/\hp\approx6\times10^{-5}$ would be 
$B({\rm 5d})P({\rm 5d,3p\ ^3P})P({\rm 3p\ ^3P,3s\ ^3P^o}) 
/E(\hb)=7.7\times10^{-4}$ 
where we took $P({\rm 3p\ ^3P,3s\ ^3P^o})=0.36$ and 
an effective recombination coefficient of the 
\term{4f\ F} term $\alpha^{\rm eff}_{\rm 4f\,F}=
1.0\times 10^{-13}\cccs$ at $10^4\K$ \citepalias{escvict}. 
The strongest line of the multiplet, 
\mult{3s\ ^3P_2^o}{3p\ ^3P_2}\lam4630.54, has an 
LS line fraction of $11.25/27$ \citep{allen} and 
the corresponding increase in intensity 
would be at least 0.034 ($I(\hb)=100$), 
which is comparable the observed intensity of 0.048 
\citepalias{esteban04} produced by more direct cascade routes 
following absorption of photons 
at \lala530 and 534\amst\ by 3d states. 
Thus the excitation of 4f states by fluorescence 
would produce lines from 3p and 3d states with intensities 
much higher than the observed values unless the stellar continuum 
had an unusual shape that selectively excited states above the 4f states. 
Our calculations show negligible 
contribution of fluorescence to the excitation of 
the 4f levels because the absorption rate is much less for 
higher resonant levels than for the 3d levels, 
and point to other mechanisms to excite them \citep{tsamis2}. 

\citetalias{esteban04} also observed the lines  
at \lam5001.14 and 5001.48 with upper levels 
\term{3d\ ^3F^o_{2,3}}. 
The most intense component of the multiplet at \lam5005.15 is 
blended with the [\oiii] line. 
As with the lines with upper 4f levels, 
the \lam5005.15 line is mostly excited by recombination 
because its upper level \term{3d\ ^3F^o_4} can receive only 
indirect contributions from the fluorescence excitation of higher levels. 
The other levels, \term{3d\ ^3F^o_{2,3}}, are connected to the ground state 
through weak dipole--allowed transitions \citep{bell}, 
and have a substantial fluorescence contribution. 
Our model calculations show that fluorescence contributes less 
than 5\% to the intensity of the lines produced by 
the \term{3d\ ^3F^o_4} and 4f levels in Orion, 
and it is not sufficient to explain the discrepancy between 
the abundances determined from recombination and collisionally 
excited lines. 

\subsection{The fluorescence spectrum}
\label{fluorspec}

%         Basic model intensities (sed370.cl.tex) 
\newcommand{\header}{%
   \multicolumn{1}{c}{Transition}
    &$\lam$(\AA)&$I_r/I$&\multicolumn{3}{c}{$I/I(\hb)$}\\%
   &&\multicolumn{1}{c}{(1)}&\multicolumn{1}{c}{(2)}%
    &\multicolumn{1}{c}{(3)}&\multicolumn{1}{c}{(4)}\\}
\begin{table}
  \caption{Predicted intensities of the basic model (BM) and observations 
($I(\hb)=10^4$) for lines excited by fluorescence. The nebular model 
was calculated with \clo.}
  \label{tabbscmod}
  \begin{tabular}{@{}rrcrll}
    \hline
    \header
    \hline
% generated with convtex.sm 
\mult{2p^3\ ^3P^o_2}{4p\ ^3S_1}& 1060.2&0.362& 0.38&-- &--  \\ 
\mult{2p^3\ ^3P^o_1}{4p\ ^3S_1}& 1060.2&0.362& 0.37&-- &--  \\ 
\mult{2p^3\ ^3P^o_0}{4p\ ^3S_1}& 1060.3&0.362& 0.15&-- &--  \\ 
\mult{2p^3\ ^3D^o_3}{3p\ ^3P_2}& 1275.0&0.118&17.31&-- &--  \\ 
\mult{2p^3\ ^3D^o_2}{3p\ ^3P_2}& 1275.3&0.118& 3.15&-- &--  \\ 
\mult{2p^3\ ^3D^o_1}{3p\ ^3P_2}& 1275.3&0.118& 0.21&-- &--  \\ 
\mult{2p^3\ ^3D^o_1}{3p\ ^3P_1}& 1276.2&0.094& 3.92&-- &--  \\ 
\mult{2p^3\ ^3D^o_2}{3p\ ^3P_1}& 1276.2&0.094&11.62&-- &--  \\ 
\mult{2p^3\ ^3D^o_1}{3p\ ^3P_0}& 1276.8&0.079& 6.01&-- &--  \\ 
\mult{2p^3\ ^3D^o_3}{3p\ ^3D_3}& 1343.3&0.257& 3.30&-- &--  \\ 
\mult{2p^3\ ^3D^o_2}{3p\ ^3D_3}& 1343.6&0.257& 0.39&-- &--  \\ 
\mult{2p^3\ ^3D^o_3}{3p\ ^3D_2}& 1345.1&0.176& 0.68&-- &--  \\ 
\mult{2p^3\ ^3D^o_1}{3p\ ^3D_2}& 1345.3&0.176& 0.55&-- &--  \\ 
\mult{2p^3\ ^3D^o_2}{3p\ ^3D_2}& 1345.3&0.176& 2.63&-- &--  \\ 
\mult{2p^3\ ^3D^o_1}{3p\ ^3D_1}& 1346.4&0.167& 1.80&-- &--  \\ 
\mult{2p^3\ ^3D^o_2}{3p\ ^3D_1}& 1346.4&0.167& 0.66&-- &--  \\ 
\mult{2p^3\ ^3P^o_1}{3p\ ^3P_2}& 1627.3&0.118& 0.13&-- &--  \\ 
\mult{2p^3\ ^3P^o_2}{3p\ ^3P_2}& 1627.4&0.118& 0.46&-- &--  \\ 
\mult{2p^3\ ^3P^o_2}{3p\ ^3P_1}& 1628.9&0.094& 0.30&-- &--  \\ 
\mult{2p^3\ ^3P^o_1}{3p\ ^3P_1}& 1628.9&0.094& 0.07&-- &--  \\ 
\mult{2p^3\ ^3P^o_0}{3p\ ^3P_1}& 1629.1&0.094& 0.11&-- &--  \\ 
\mult{2p^3\ ^3P^o_1}{3p\ ^3P_0}& 1629.8&0.079& 0.18&-- &--  \\ 
\mult{2p^3\ ^3P^o_1}{3p\ ^3S_1}& 1675.7&0.084& 5.43&-- &--  \\ 
\mult{2p^3\ ^3P^o_2}{3p\ ^3S_1}& 1675.8&0.084& 8.87&-- &--  \\ 
\mult{2p^3\ ^3P^o_0}{3p\ ^3S_1}& 1675.9&0.084& 1.83&-- &--  \\ 
\mult{2p^3\ ^3P^o_2}{3p\ ^3D_3}& 1740.3&0.257& 9.39&-- &--  \\ 
\mult{2p^3\ ^3P^o_2}{3p\ ^3D_2}& 1743.2&0.176& 2.39&-- &--  \\ 
\mult{2p^3\ ^3P^o_1}{3p\ ^3D_2}& 1743.2&0.176& 7.32&-- &--  \\ 
\mult{2p^3\ ^3P^o_1}{3p\ ^3D_1}& 1745.0&0.167& 2.54&-- &--  \\ 
\mult{2p^3\ ^3P^o_2}{3p\ ^3D_1}& 1745.1&0.167& 0.17&-- &--  \\ 
\mult{2p^3\ ^3P^o_0}{3p\ ^3D_1}& 1745.3&0.167& 3.45&-- &--  \\ 
\mult{3s\ ^3P^o_0}{4p\ ^3S_1}& 1830.5&0.362& 0.01&-- &--  \\ 
\mult{3s\ ^3P^o_1}{4p\ ^3S_1}& 1831.6&0.362& 0.02&-- &--  \\ 
\mult{3s\ ^3P^o_2}{4p\ ^3S_1}& 1836.2&0.362& 0.06&-- &--  \\ 
\mult{2p^3\ ^1D^o_2}{3p\ ^1D_2}& 3329.7&0.454& 0.01&-- &--  \\ 
\mult{3s\ ^3P^o_1}{3p\ ^1D_2}& 3955.8&0.454& 0.06&-- &--  \\ 
\mult{3s\ ^1P^o_1}{3p\ ^1D_2}& 3995.0&0.454& 0.60&-- &1.0? \\ 
\mult{3p\ ^1P_1}{3d\ ^3P^o_1}& 4114.3&0.014& 0.01&-- &--  \\ 
\mult{3p\ ^1P_1}{3d\ ^3D^o_2}& 4375.0&0.035& 0.03&-- &--  \\ 
\mult{3p\ ^1P_1}{3d\ ^3D^o_1}& 4379.6&0.016& 0.01&-- &--  \\ 
\mult{3p\ ^3D_1}{3d\ ^3P^o_0}& 4459.9&0.014& 0.15&-- &--  \\ 
\mult{3p\ ^3D_1}{3d\ ^3P^o_1}& 4465.5&0.014& 0.09&1.5?&--  \\ 
\mult{3p\ ^3D_2}{3d\ ^3P^o_1}& 4477.7&0.014& 0.34&-- &--  \\ 
\mult{3p\ ^3D_2}{3d\ ^3P^o_2}& 4488.1&0.033& 0.06&-- &--  \\ 
\mult{3p\ ^3D_3}{3d\ ^3P^o_2}& 4507.6&0.033& 0.45&-- &--  \\ 
\mult{3p\ ^1P_1}{3d\ ^3F^o_2}& 4564.8&0.467& 0.02&-- &--  \\ 
\mult{3s\ ^3P^o_1}{3p\ ^3P_2}& 4601.5&0.118& 2.01&1.5 &1.3  \\ 
\mult{3s\ ^3P^o_0}{3p\ ^3P_1}& 4607.2&0.094& 2.11&5.7?&4.2  \\ 
\mult{3s\ ^3P^o_1}{3p\ ^3P_1}& 4613.9&0.094& 1.46&0.9?&1.0  \\ 
\mult{3s\ ^3P^o_1}{3p\ ^3P_0}& 4621.4&0.079& 2.38&1.8?&1.6  \\ 
\mult{3s\ ^3P^o_2}{3p\ ^3P_2}& 4630.5&0.118& 6.57&4.6 &4.8  \\ 
\mult{3s\ ^3P^o_2}{3p\ ^3P_1}& 4643.1&0.094& 2.89&2.2 &1.5  \\ 
\mult{3s\ ^1P^o_1}{3p\ ^3P_2}& 4654.5&0.118& 0.21&-- &--  \\ 
\mult{3s\ ^1P^o_1}{3p\ ^3P_1}& 4667.2&0.094& 0.19&-- &--  \\ 
\mult{3s\ ^1P^o_1}{3p\ ^3P_0}& 4674.9&0.079& 0.26&-- &--  \\ 
% generated with convtex.sm 
    \hline
  \end{tabular}
\end{table}
\begin{table}
  \contcaption{}
  \begin{tabular}{@{}crcrll}
    \hline
    \header
    \hline
% (continuation) generated with convtex.sm 
\mult{3p\ ^3D_1}{3d\ ^3D^o_2}& 4774.2&0.035& 0.16&-- &--  \\ 
\mult{3p\ ^3D_1}{3d\ ^3D^o_1}& 4779.7&0.016& 1.11&0.8 &1.1  \\ 
\mult{3p\ ^3D_2}{3d\ ^3D^o_3}& 4781.2&0.096& 0.09&-- &--  \\ 
\mult{3p\ ^3D_2}{3d\ ^3D^o_2}& 4788.1&0.035& 1.26&1.2 &1.4  \\ 
\mult{3p\ ^3D_2}{3d\ ^3D^o_1}& 4793.6&0.016& 0.34&1.1?&--  \\ 
\mult{3p\ ^3D_3}{3d\ ^3D^o_3}& 4803.3&0.096& 1.45&1.3 &1.9  \\ 
\mult{3p\ ^3D_3}{3d\ ^3D^o_2}& 4810.3&0.035& 0.24&-- &--  \\ 
\mult{2p^3\ ^1D^o_2}{3p\ ^1P_1}& 4895.1&0.231& 0.07&1.7?&--  \\ 
\mult{3p\ ^3S_1}{3d\ ^3P^o_0}& 4987.4&0.014& 0.89&7.6?&4.6? \\ 
\mult{3p\ ^3S_1}{3d\ ^3P^o_1}& 4994.4&0.014& 2.65&1.3?&1.8  \\ 
\mult{3p\ ^3D_1}{3d\ ^3F^o_2}& 5001.1&0.467& 1.10&-- &1.2  \\ 
\mult{3p\ ^3D_2}{3d\ ^3F^o_3}& 5001.5&0.375& 2.30&-- &1.8  \\ 
\mult{3s\ ^3P^o_0}{3p\ ^3S_1}& 5002.7&0.084& 0.53&-- &--  \\ 
\mult{3p\ ^3D_3}{3d\ ^3F^o_4}& 5005.2&0.944& 1.44&-- &--  \\ 
\mult{3p\ ^3S_1}{3d\ ^3P^o_2}& 5007.3&0.033& 3.23&-- &--  \\ 
\mult{3s\ ^3P^o_1}{3p\ ^3S_1}& 5010.6&0.084& 1.38&-- &--  \\ 
\mult{3p\ ^3D_2}{3d\ ^3F^o_2}& 5016.4&0.467& 0.18&-- &--  \\ 
\mult{3p\ ^3D_3}{3d\ ^3F^o_3}& 5025.7&0.375& 0.23&-- &--  \\ 
\mult{3s\ ^3P^o_2}{3p\ ^3S_1}& 5045.1&0.084& 2.13&-- &1.4  \\ 
\mult{3s\ ^1P^o_1}{3p\ ^3S_1}& 5073.6&0.084& 0.16&-- &--  \\ 
\mult{3p\ ^3S_1}{3d\ ^3D^o_2}& 5383.7&0.035& 0.01&-- &--  \\ 
\mult{3p\ ^3S_1}{3d\ ^3D^o_1}& 5390.7&0.016& 0.01&-- &--  \\ 
\mult{3p\ ^3P_0}{3d\ ^3P^o_1}& 5452.1&0.014& 0.28&-- &--  \\ 
\mult{3p\ ^3P_1}{3d\ ^3P^o_0}& 5454.2&0.014& 0.36&-- &--  \\ 
\mult{3p\ ^3P_1}{3d\ ^3P^o_1}& 5462.6&0.014& 0.32&-- &--  \\ 
\mult{3p\ ^3P_1}{3d\ ^3P^o_2}& 5478.1&0.033& 0.18&-- &--  \\ 
\mult{3p\ ^3P_2}{3d\ ^3P^o_1}& 5480.1&0.014& 0.41&-- &--  \\ 
\mult{3p\ ^3P_2}{3d\ ^3P^o_2}& 5495.7&0.033& 0.89&0.7 &0.5? \\ 
\mult{3s\ ^3P^o_1}{3p\ ^3D_2}& 5666.6&0.176& 4.63&3.1 &2.9  \\ 
\mult{3s\ ^3P^o_0}{3p\ ^3D_1}& 5676.0&0.167& 2.32&1.2 &1.0? \\ 
\mult{3s\ ^3P^o_2}{3p\ ^3D_3}& 5679.6&0.257& 6.24&4.3 &4.3  \\ 
\mult{3s\ ^3P^o_1}{3p\ ^3D_1}& 5686.2&0.167& 1.52&0.8 &0.6? \\ 
\mult{3s\ ^3P^o_2}{3p\ ^3D_2}& 5710.8&0.176& 1.52&0.9 &0.9  \\ 
\mult{3s\ ^3P^o_2}{3p\ ^3D_1}& 5730.7&0.167& 0.10&-- &--  \\ 
\mult{3s\ ^1P^o_1}{3p\ ^3D_2}& 5747.3&0.176& 0.41&-- &--  \\ 
\mult{3s\ ^1P^o_1}{3p\ ^3D_1}& 5767.5&0.167& 0.19&-- &--  \\ 
\mult{3p\ ^3P_0}{3d\ ^3D^o_1}& 5927.8&0.016& 1.15&0.7?&1.0? \\ 
\mult{3p\ ^3P_1}{3d\ ^3D^o_2}& 5931.8&0.035& 1.72&1.4 &2.0  \\ 
\mult{3p\ ^3P_1}{3d\ ^3D^o_1}& 5940.2&0.016& 0.80&-- &--  \\ 
\mult{3p\ ^3P_2}{3d\ ^3D^o_3}& 5941.7&0.096& 2.05&1.2 &1.5  \\ 
\mult{3p\ ^3P_2}{3d\ ^3D^o_2}& 5952.4&0.035& 0.51&0.6?&1.2? \\ 
\mult{3p\ ^3P_2}{3d\ ^3D^o_1}& 5960.9&0.016& 0.05&-- &--  \\ 
\mult{3s\ ^3P^o_1}{3p\ ^1P_1}& 6379.6&0.231& 0.08&-- &--  \\ 
\mult{3s\ ^1P^o_1}{3p\ ^1P_1}& 6482.0&0.231& 0.38&-- &--  \\ 
\mult{3d\ ^3P^o_2}{4p\ ^3S_1}& 6810.0&0.362& 0.04&-- &0.3? \\ 
\mult{3d\ ^3P^o_1}{4p\ ^3S_1}& 6834.1&0.362& 0.02&-- &--  \\ 
\mult{3d\ ^3P^o_0}{4p\ ^3S_1}& 6847.2&0.362& 0.01&-- &--  \\ 
\mult{4s\ ^3P^o_2}{4p\ ^3S_1}&14645.6&0.362& 0.01&-- &--  \\ 
% (continuation) generated with c^onvtex.sm 
    \hline
  \end{tabular} 

  \medskip
  (1) Recombination contribution to total intensity.\\
  (2) Predicted intensities with an \hb\ flux of 2.51\flu\\
  (3) \citetalias{baldwin00}\\
  (4) \citetalias{esteban04}\\
\end{table}

Table~\ref{tabbscmod} gives the predicted intensities for the 
observed upper terms in Orion by \citetalias{baldwin00} and 
\citetalias{esteban04} that are excited mostly by fluorescence. 
Although the two data sets are from different parts of the 
nebula, it is important to notice that both sets give 
similar measurements of the \nii\ permitted line intensities with 
respect to \hb. 
Uncertain observed intensities due to blends, 
low S/N or dubious identifications are marked with ``?'' 
in the table as indicated by those authors. 
We also added a question mark to the line at 4987.4\amst, 
which is probably blended with the 
[\mbox{Fe\thinspace{\sc III}}]\lam4987.20 line,  
and thus has an observed intensity much higher than our prediction. 
The line at 4994.4\amst\ belonging to the same multiplet should 
be theoretically more intense, contrary to the observations. 

The most intense 
fluorescence lines are triplets connected by resonant 
transitions to the ground term, \term{2p^2\ ^3P}. 
At temperatures characteristic of \hii\ regions, 
the fine structure populations of the ground term are 
approximately proportional to their statistical weight, 
and consequently the relative intensities of all the 
other triplet levels are given by the LS line fractions \citep{allen}. 

The lines at 5001.15 and 5001.48\amst\ 
arising from the \term{3d\ ^3F^o} term are blended. 
We split the total 
intensity according to the LS line fractions: 21.0:31.1, in order 
to compare them with our predictions in table~\ref{tabbscmod}. 

Multiplet \mult{3p\ ^3P}{4s\ ^3P^o} has many 
strong lines in the \lam3838.37--3856.06\amst\ interval that 
were not detected because of the lower S/N at the 
blue end of the spectrum and because they are blended with other lines. 
Therefore they are not listed in table~\ref{tabbscmod}. 

\citetalias{esteban04} marginally detected the 
\mult{3d\ ^3P^o_2}{4p\ ^3S_1}\lam6809.99\amst\ line. 
Lines from the 4p levels must be excited by cascades 
from upper levels in a fluorescence spectrum. 
We found very low intensities for all the lines from 4p levels. 
The most intense line of this type is 
\mult{2p^3\ ^3P_2^o}{4p\ ^3S_1}\lam1060.2\amst\ with an 
intensity of $4\times10^{-5}$ with respect to \hb.
Our predicted intensities of lines from 4p levels in 
the optical are below the instrumental sensitivity. 

In table~\ref{tabbscmod} we also list some singlets because 
\citetalias{baldwin00} marginally detected the 
\mult{2p^3\ ^1D^o_2}{3p\ ^1P_1}\lam4895.11\amst\ line. 
Singlets can be excited by absorptions from the 
\term{^1D} and \term{^1S} terms of the 
ground configuration, spin--forbidden transitions from the 
triplets and recombination. 
The quintet lines 
\mult{3p\ ^5D^o_1}{3d\ ^5P_1}\lam4815.62\amst, 
\mult{3s\ ^5P_{1,3}}{3p\ ^5D_{1,4}}\lam5535.36\amst\ and 
\mult{3s\ ^5P_3}{3p\ ^5D_3}\lam5551.99\amst\ measured 
by \citetalias{baldwin00} and \citetalias{esteban04} are not listed. 
These lines have the \term{2s2p^2\ ^4P} excited core configuration, 
and can only be populated with transitions in which the 
LS coupling breaks down. 
The \term{ 3d\ ^5P} term is autoionizing, 
and the most likely mechanism in this case is dielectronic recombination. 
\citetalias{esteban04} reported lines at \lala6744.42, 7535.32 and 
9016.42\amst\ that would be produced by highly excited d and f 
states of \np. 
Their identification as \nii, however, is uncertain 
and their production in our model is very unlikely. 
Many more lines are predicted by our calculations, 
but their intensities are lower than the weaker lines reported 
by \citetalias{baldwin00} or \citetalias{esteban04}, 
and their intensities are dominated by recombination. 
Intensities for other lines up to $n=8$ and $l=1$ are 
available from the authors upon request. 
Most of the observed lines with confident measurements 
fall within 0.2 dex of the predicted values in the basic model 
as shown in the upper panels of Fig.~\ref{nebuvsobs}. 

\section{Variation of parameters}

\subsection{Stellar temperature}
\label{stellartemp}

%                                         generated with comps.sm 
\begin{figure}
 \includegraphics[width=8.8cm]{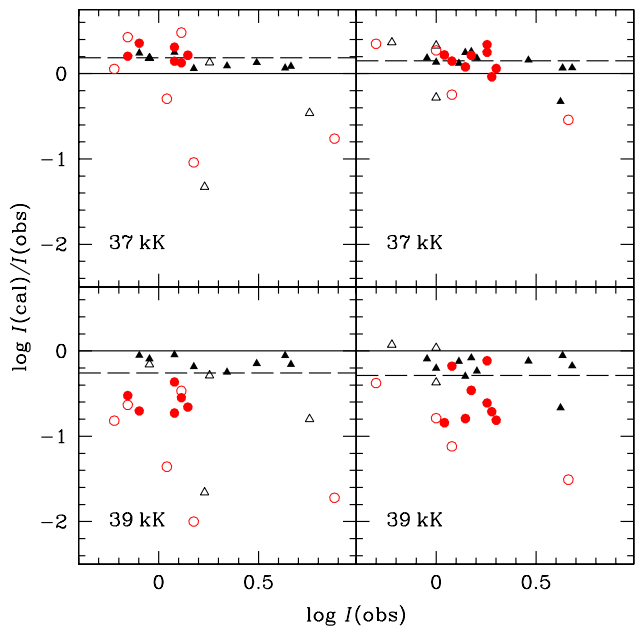}
 \caption{Comparison of intensities observed by \citetalias{baldwin00} 
  (left) and \citetalias{esteban04} (right panels) with predicted 
  intensities of fluorescence lines from p states (triangles) 
  and d states (circles) for two WMbasic stellar atmospheres. 
  Open triangles and circles have uncertain $I_{\rm obs}$ 
  (marked with ``?'' in table~\ref{tabbscmod}). 
  The broken horizontal line is the value of 
  $R=\langle I_{\rm cal}/I_{\rm obs}\rangle$. 
  The nebular model was calculated with \nebu\ with BM parameters,  
  except for $\teff=39\kK$ in the lower panels, giving column densities 
  of $N(\np)=7.3\times10^{16}\persqcm$ and $2.6\times10^{16}\persqcm$ 
  for the $\teff=37\kK$ and $39\kK$ atmospheres respectively. Note 
  that the observed intensities here and in table~\ref{tabbscmod} 
  are normalized to $I(\hb)=10^4$.} 
 \label{nebuvsobs}
\end{figure}

%                             generated with steflu.sm
\begin{figure}
 \includegraphics[width=8.8cm]{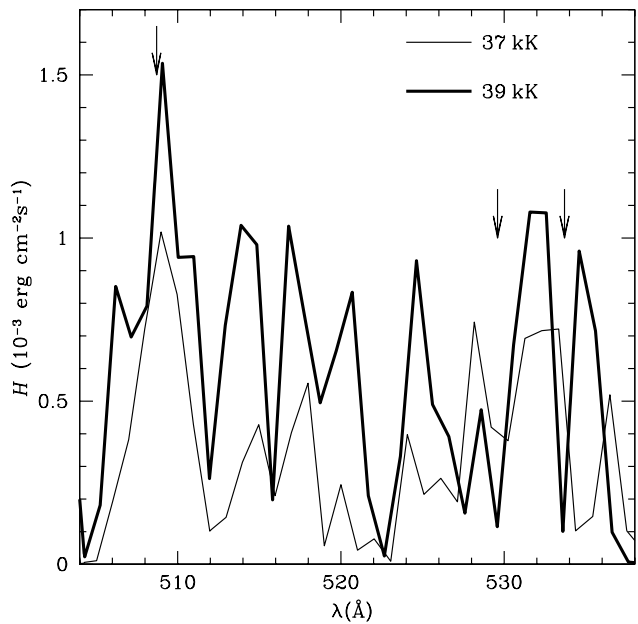}
 \caption{Surface Eddington flux of two WMbasic atmospheres. 
  Arrows show wavelengths of the resonant multiplets 
  at 508.7, 529.6 and 533.7\amst.}
 \label{fvslam}
\end{figure}

%                             generated with inttef.sm
\begin{figure}
 \includegraphics[width=8.8cm]{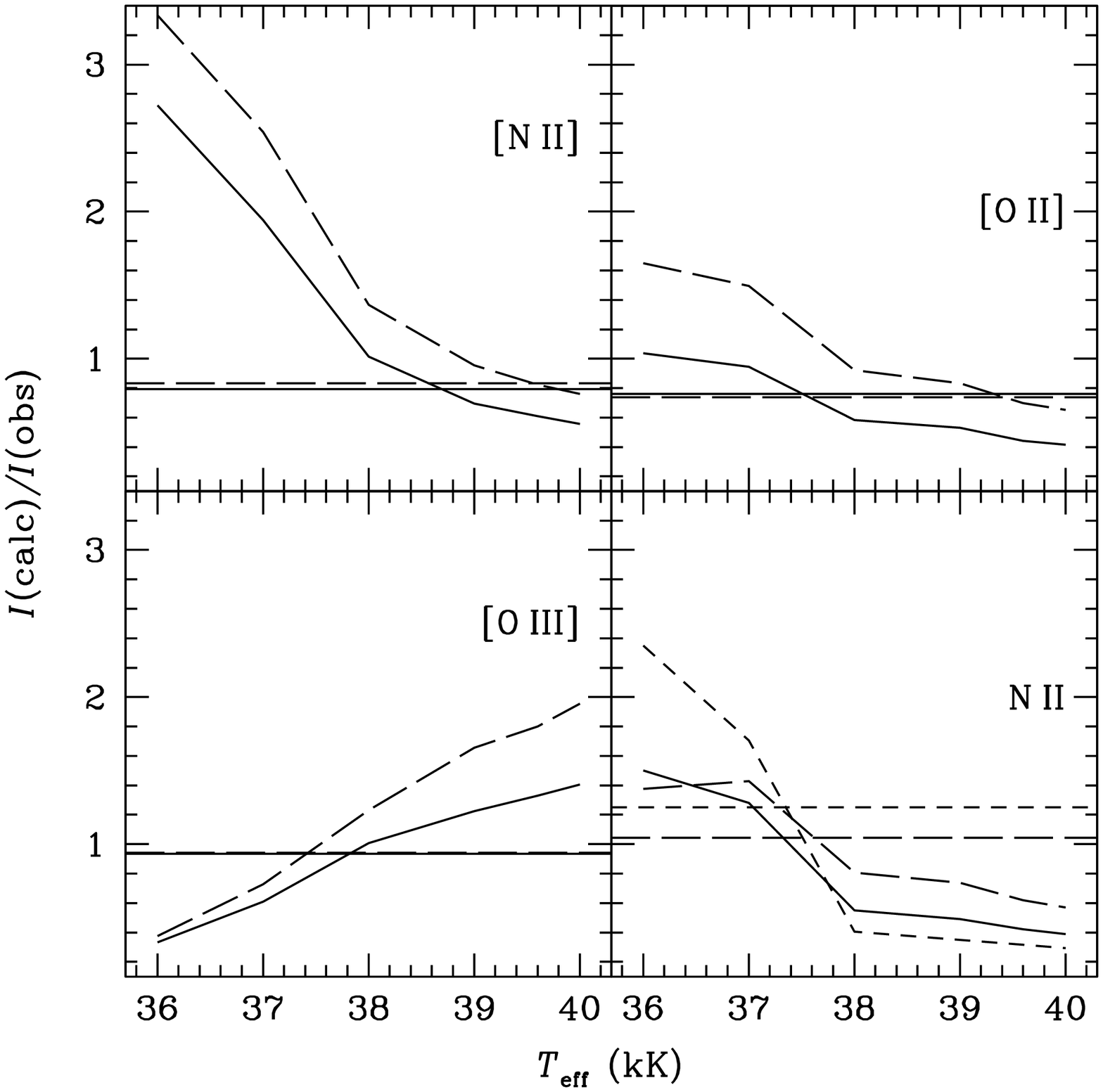}
 \caption{Intensities of some nebular diagnostic and 
  \nii\ permitted lines normalized to the observed values 
  by \citetalias{baldwin00} as a function of stellar \teff. 
  Top left: [NII] 6584 (solid), [NII]5755 (dashed). 
  Top right: [OII] 3729 (solid), [OII] 3726 (dashed). 
  Bottom left: [OIII] 5007 (solid), [OIII] 4363 (dashed). 
  Bottom right: \nii\ permitted lines averaged 
  $R=\langle I_{\rm cal}/I_{\rm obs}\rangle$ (solid), 
  \lam5941.65\amst (d upper state, short dash) 
  and 4630.54\amst\ (p upper state, long dash). 
  Horizontal lines are the values observed 
  by \citetalias{esteban04}. 
  The nebular model was calculated with \clo\ with the BM parameters,  
  except for a varying \teff.} 
 \label{ivsteff}
\end{figure}

Fig.~\ref{nebuvsobs} shows a comparison of 
observed and predicted intensities with two WMbasic atmospheres. 
The reduction in \np\ column density produced by the hotter 
star reduces the predicted intensities in general, 
but the intensities of the lines originating 
from d states suffer a much greater reduction than the ones from p states
by an order of magnitude. 
We have traced this effect to an important 
decrease in the model atmosphere flux by a factor of $\sim2$ at 
529.6\amst\ and a factor of $\sim5$ at 533.7\amst\ for 
$\teff\ga38\kK$ as shown in Fig.~\ref{fvslam}. 
As mentioned in section~\ref{atprocs}, 
3d states are pumped almost entirely by 
absorptions at those two wavelengths. 
At the same time 
there is an increase of a factor of $\sim2$ in the flux at 
508.7\AA\ for $\teff\ga37\kK$, 
which is important in the pumping of 3p states, 
and compensates the decrease in the \np\ column density.  

The difference in the pumping rates of d and p states 
for $\teff\ga38\kK$ depends only on the shape of the spectrum 
and the contribution of recombination to the population 
of those states. 
Thus the disagreement between predicted and observed 
intensities of lines from d states with $\teff=39\kK$ 
shown in Fig.~\ref{nebuvsobs} 
persists when the N abundance is 
increased to $1\times10^{-4}$ or when the 
recombination contribution to the intensities is 
increased with the hypothesis of ultracold plasma 
proposed by \citet{tsamis2}. 
A plasma temperature of $2000\K$ doubles the predicted intensities 
of lines from 3p states,  
but lines from 3d states increase their intensities in 
lower proportions because fluorescence is more important 
in the population of those states. 

The other critical parameter in the fluorescence 
line intensities is the column density. 
The \np\ column density 
decreases with \teff\ at a much lower rate for 
$\teff\ga38\kK$ as reflected in Fig.~\ref{ivsteff}. 
The NII fluorescence lines and the [\nii] lines 
decrease little for harder spectra due to that persistent 
\np\ concentration, 
but their different behaviour at lower \teff\ can be 
understood in terms of the \np\ concentration and the 
escape probability concept. 
As shown in Fig.~\ref{emisvsr}, 
the fluorescence \nii\ lines 
form 50\% of their intensity in the inner layers of the 
nebula, much closer to the star than the [\nii] lines, 
which are produced in the outer \np\ zone. 
As \teff\ decreases, 
the \np\ concentration and the 
optical depth of the resonant transitions increase, 
but the intensities of lines from p and d states behave 
differently as shown in Fig.~\ref{ivsteff} for the 
lines \mult{3s\ ^3P_2^o}{3p\ ^3P_2}\lam4630.54 and 
\mult{3p\ ^3P_2}{3d\ ^3D_3^o}\lam5941.65. 
Absorption transitions that populate 
the p states have a much lower optical depth than the ones 
pumping the d states. 
As a result the escape probability decreases more for d states 
than for p states with lower \teff, 
and the pumping due to reabsorption of resonant photons for 
d states increases. 

\nebu\ tends to give larger column densities than \clo, 
and thus predicts higher intensities. 
Predicted line intensities by \nebu\ in the UV tend to be 30\% more 
intense than those given by \clo\ because \nebu\ does not 
consider internal dust extinction, but predictions of the 
two codes are within 20\% of each other in the optical. 

\subsection{Kurucz atmospheres}

\begin{figure}
 \includegraphics[width=8.8cm]{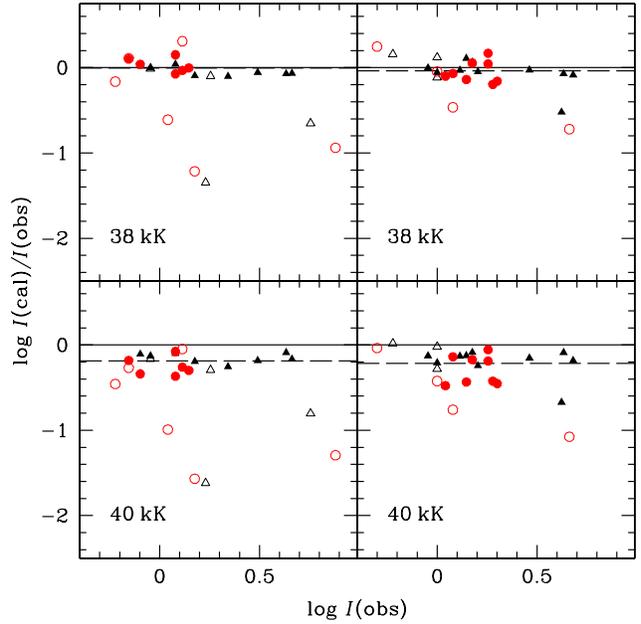}
 \caption{Same as Fig.~\ref{nebuvsobs}, but for two Kurucz 
  atmospheres. The \np\ column densities are 
  $N(\np)=3.5\times10^{16}\persqcm$ and $1.13\times10^{16}\persqcm$ 
  for the $\teff=38\kK$ and $40\kK$ atmospheres respectively.} 
 \label{kuruvsobs}
\end{figure}

\clo\ contains a grid of low--resolution Kurucz atmospheres \citep{kuru} 
that can be readily used as continua in our calculations. 
A comparison of Fig.~\ref{nebuvsobs} and~\ref{kuruvsobs} 
shows that the differences between the calculated intensities of 
p and d states with Kurucz atmospheres are much smaller than 
the differences with the WMbasic atmospheres because the 
Kurucz atmospheres do not have the structure of the WMbasic 
atmospheres that causes the different absorption rates between 
p and d states. 

Modeling of Orion with \clo\ \citep{baldwin91,baldwin96,baldwin00} 
has favored stellar temperatures that are lower than current 
spectro--photometric measurements. 
Our results with the WMbasic atmospheres also favor a lower \teff, 
but Kurucz atmospheres give a better agreement with observations 
because they are softer than other models and give a larger 
\np\ column. 

%                                         generated with comps.sm 

\subsection{Stellar flux}

%                             generated with intphi.sm 
\begin{figure}
 \includegraphics[width=8.8cm]{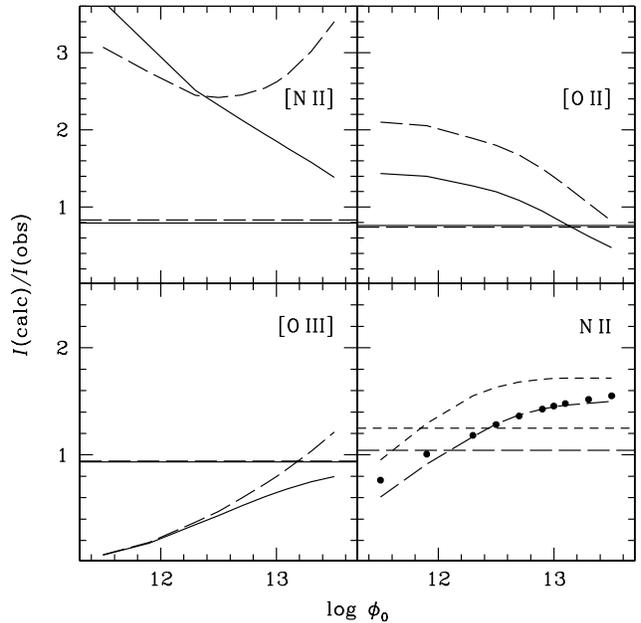}
 \caption{Same as Fig.~\ref{ivsteff}, 
  but with \teff=37kK and varying $\phi_0$. 
  The average $R=I_{\rm cal}/I_{\rm obs}$ (not shown) is within 
  3\% of the 4630.54\amst\ line. 
  The fit of \ecua{fit} (dots) is normalized to the 4630.54\amst\ line at 
  $\log\phi_0=12.9$. 
  } 
 \label{ivsphi}
\end{figure}

Unlike most forbidden and recombination lines,  
fluorescence line intensities are more sensitive 
to changes in the stellar flux illuminating the nebula. 
Their intensities increase with $\phi_0$ 
up to $10^{12.5}\persqcm\persecond$ 
and remain nearly constant for higher $\phi_0$. 
This behavior can be understood in similar terms to 
the curve of growth of the resonant lines. 
As the intensity of the ionizing flux grows, 
the \np\ column density and the optical depth increase, 
and the cores of the resonant lines become saturated. 
Eqs.~\ref{beta} and~\ref{intens} show that 
the intensity of the fluorescence lines is approximately proportional 
to the integral along the line of sight of the pumping rate of 
\ecua{beta} times the density of the absorbing state, $n_g\beta_{gj}$. 
Changing variable from $r$ to $\tau_0$, 
eliminating constant quantities 
and assuming a constant Doppler width, 
the intensity of a fluorescence line will be 
\begin{eqnarray*}
  I\propto\int_0^\infty \bar J_\nu
  \int e^{-\tau_0\phi(x)}\phi(x)\,dx\,d\tau_0 \\
 =J_\nu(0)\int_0^\infty e^{-\tau_c} 
  \int e^{-\tau_0\phi(x)}\phi(x)\,dx\,d\tau_0 \ .
\end{eqnarray*}
where $J_\nu(0)$ is the stellar continuum at the 
illuminated face of the cloud, 
and $\tau_c$ is the continuum opacity. 
The integration over $\tau_0$ can be performed exactly 
if we assume a mean value for $e^{-\tau_c}$. 
For fixed \teff\ and $\nu$, 
$J_\nu(0)$ is proportional to $\phi_0$, which in turn is 
proportional to the \hb\ flux. 
Therefore the fluorescence line intensity normalized to \hb\ must 
be proportional to 
\[
   \left< e^{-\tau_c}\right>\int(1-e^{-\tau_0\phi(x)})\,dx \ .
\]
The integral is proportional to the curve of growth $W(\tau_0)$. 
Fig.~\ref{ivsphi} shows 
that the intensity of the lines follows closely a fit of the form 
\beq
   I/I(\hb)\propto e^{-2.5\times10^{-22}N(\hp)}W(\tau_0) \ .
   \label{fit}
\eeq
where $N(\hp)$ is the \hp\ column density in \persqcm\  
and $W(\tau_0)$ is the curve of growth of the 
\mult{2p^2\ ^3P_2}{4s\ ^3P^o_2}\lam508.697\amst\ line, 
which pumps most of the 4630.54 line. 

\section{Conclusions}

The intensity of the lines in the \nii\ spectrum of the 
Orion nebula can be explained 
by fluorescence of the UV radiation 
of \oric\ in the ionized gas. 
Recombination of \npp\ contributes a minor 
part of the observed intensities of lines from 3p and 3d levels 
connected to the ground state. 
The effective temperature of the star must be 
below 38000\K\ in order to reproduce the observed 
line intensities with typical ionization models that 
are consistent with the forbidden line intensities. 
An increased N abundance does not allow the 
use of a higher star temperature. 
The existence of intervening ionized material in 
the foreground \citep{odell} was not considered in our model and 
may help increase the predicted intensities of the 
\nii\ lines. 
Fluorescence does not increase the intensity of the lines 
from 4f levels, 
and other mechanisms must be proposed to 
explain their strong intensities with respect 
to the collisionally excited and fluorescence lines in 
the Orion nebula.  

\section*{Acknowlegments} The authors are very grateful to Katia Verner 
and Gary Ferland for valuable advice in running \clo.

\label{lastpage}

\end{document}